\documentclass[%
 preprint,
 amsmath,amssymb,
 aps,
prl,
]{revtex4-2}

\usepackage{graphicx}
\usepackage{dcolumn}
\usepackage{bm}
\usepackage{tikz}
\usepackage{braket}
\usepackage{siunitx}
\usepackage{natbib}
\usepackage[version=4]{mhchem}

\usepackage{hyperref}

\begin{document}


\title{Equivariant Neural Networks for Spin Dynamics Simulations of Itinerant Magnets}

\author{Yu Miyazaki}
\email[]{yumiyazaki@g.ecc.u-tokyo.ac.jp}
\affiliation{Department of Applied Physics, The University of Tokyo, Hongo, Bunkyo, Tokyo 113-8656, Japan}

\date{\today}

\begin{abstract}  

I present a novel equivariant neural network architecture for the large-scale spin dynamics simulation of the Kondo lattice model. 
This neural network mainly consists of tensor-product-based convolution layers  and ensures two equivariances: translations of the lattice and rotations of the spins.
I implement equivariant neural networks for two Kondo lattice models on two-dimensional square and triangular lattices, and perform training and validation. In the equivariant model for the square lattice, the validation error (based on root mean squared error) is reduced to less than one-third compared to a model using invariant descriptors as inputs. Furthermore, I demonstrate the ability to reproduce  phase transitions of skyrmion crystals in the triangular lattice, by performing dynamics simulations using the trained model.
\end{abstract}

\maketitle
\section{Introduction}
While localized magnetism has been studied with great success for many years, itinerant magnetism is still only partially understood \cite{Santiago2017}.
Many magnetic materials are in the intermediate region between itinerant magnetism and localized magnetism, and the elucidation of itinerant magnetism is still an important research objective.
The Ruderman-Kittel-Kasuya-Yosida (RKKY) interaction is known as the lowest-order interaction between two localized spins mediated by itinerant electrons. In recent years, it is believed that more complex spin interactions originating from itinerant electrons, which cannot be explained by a simple RKKY mechanism, play an important role, particularly in the context of searching for skyrmions in centrosymmetric systems \cite{Kurumaji2019,Hirschberger2020,Hirschberger2020_2,Nomoto2020,Hirschberger2021,Khanh2020,Yasui2020,Hayami2021square}. In fact, it has been suggested that the four-spin interaction mediated by itinerant electrons may be crucial for the formation of skyrmions in  centrosymmetric square systems, \ce{GdRu2Si2} \cite{Khanh2020,Yasui2020,Hayami2021square} and \ce{EuAl4} \cite{Takagi2022}.

Despite the long history of research on d- or f-magnetic metals, direct simulations of the Kondo lattice model, which involves the interaction between localized spins and itinerant electrons, have not been conducted until recently \cite{Barros2013,Hayami2021locking,Hayami2021temperature,Eto2022,Ozawa2017}. To handle itinerant electrons, diagonalization of matrices is required; however, the computational cost of standard diagonalization is $\mathcal{O}(N^3)$, making large-scale calculations difficult. Linear-scaling approaches, such as the kernel polynomial
method (KPM) \cite{Barros2013,Wang2018} have been proposed and are often used, but it requires large-scale parallelization, and the overhead of parallelization cannot be ignored.

Recently, surrogate models to speed up and scale up simulations by replacing the calculation of energy by exact diagonalization (ED) with neural networks (NNs) have appeared \cite{Zhang2021}.
This framework is similar to machine learning interatomic potentials, which allow for large-scale molecular dynamics simulations with the accuracy of density-functional theory \cite{Behler2007,Bartok2010,Behler2016,Smith2017,Takamoto2022}.
In the previous study \cite{Zhang2021}, they use $SO(3)$-invariant descriptor (dot products $b_{jk}=\bm{S}_{\bm{r}_j}\cdot\bm{S}_{\bm{r}_k}$ and scalar triple products $\chi_{jmn} = \bm{S}_{\bm{r}_j}\cdot(\bm{S}_{\bm{r}_m}\times \bm{S}_{\bm{r}_n})$) as inputs and ensure that the energy of itinerant electrons $E_{el}$ is invariant for rotations of spins.
This corresponds to traditional Böhler-Parinello type machine learning interatomic potentials \cite{Behler2007}, which use invariant descriptors as input.

In recent years, the concept of equivariance has gained attention in deep learning, and its practical applications have advanced in fields such as materials science and computer vision \cite{Cohen2016,Thomas2018,Batzner2022,Cohen2018,Cohen2017,Oliver2021,David2020}.
Equivariance is a property of certain functions or models that ensures that their output changes consistently with respect to transformations applied to their input. In other words, when a transformation, such as rotation or translation, is applied to the input, the output also undergoes a similar transformation. This property is particularly useful in various fields, including physics and computer vision, as it allows models to maintain a coherent relationship between input and output data despite changes in the input's structure or orientation.

In this paper, I present an equivariant convolutional neural network (ECNN) architecture as a surrogate model to the calculation of itinerant electrons in the Kondo lattice model. 
I focus on two operations: translations of the lattice and rotations of the spins. Convolutional neural networks (CNNs) are equivariant with respect to the translation operation \cite{Cohen2016}, and the tensor product expansion using Clebsch-Gordan coefficients is equivariant with respect to the rotation operation \cite{Thomas2018,Kondor2018,Oliver2021}. 
By integrating these architectures, I define a convolutional layer that is equivariant to both of these operations.
For the case of the square lattice, this ECNN demonstrates superior predictive performance, with the validation error based on root mean square error approximately 1/3 that of a fully connected neural network using  invariant descriptors. 
Moreover, for the case of the triangular lattice, I conduct dynamics simulations and confirm the ability to reproduce the phase transitions of the skyrmion lattices.

\section{Result}

\subsection{Kondo Lattice Model}
I consider the Kondo lattice model (double exchange model) consisting of itinerant electrons and localized spins, whose Hamiltonian is given by
\begin{align}
    \mathcal{H}(\mathcal{S}) &= \mathcal{H}_{el}(\mathcal{S}) + \mathcal{H}_{s}(\mathcal{S}) \label{eq:H} \\ 
    \mathcal{H}_{el}(\mathcal{S}) &= -\sum_{\bm{r}\bm{r}'\alpha}t_{\bm{r}\bm{r}'}c^\dagger_{\bm{r}\alpha}c_{\bm{r}'\alpha} + \mathrm{c.c.} -J\sum_i\bm{s}_{\bm{r}}\cdot\bm{S}_{\bm{r}} \label{eq:H_el} \\ 
    \mathcal{H}_{s}(\mathcal{S}) &= - H_z \sum_{\bm{r}}S_{\bm{r}}^z. \label{eq:H_s}
\end{align}
Here, $\bm{S}_{\bm{r}}$ are (classical) localized spins with a fixed length $|\bm{S}_i|=1$ and $\mathcal{S}=\left\{ \bm{S}_{\bm{r}}\right\}$ (for the lattice size $N$) is a set of whole localized spins, $c^\dagger_{\bm{r}\alpha}$ ($c_{\bm{r}\alpha}$) are  creation (annihilation) operators of an itinerant electron at site $\bm{r}$ with spin $\alpha$, $\bm{s}_{\bm{r}} = (1/2)\sum_{\alpha\beta}c^\dagger_{\bm{r}\alpha}\bm{\sigma}_{\alpha\beta}c_{\bm{r}\beta}$ are itinerant electron spins, and $\bm{\sigma}= (\sigma^x,\sigma^y,\sigma^z)$ is a vector of Pauli matrices.
The first term of eq.~\eqref{eq:H_el} represents the kinetic energy of itinerant electrons, the second term of eq.~\eqref{eq:H_el} represents the Hund coupling between itinerant electron spins $\bm{s}_i$ and localized spins $\bm{S}_i$.
$\mathcal{H}_s$ in eq.~\eqref{eq:H_s} induludes only classical spin terms. In this paper, I only consider the Zeeman coupling to an external magnetic field $H_z$ along the $z$-direction. 
However, I note that this term does not affect the input/output relations of the neural networks and can include any localized-spin-only term.

\subsection{Equivariance}
\begin{figure*}
    \centering
    \includegraphics[width=\textwidth]{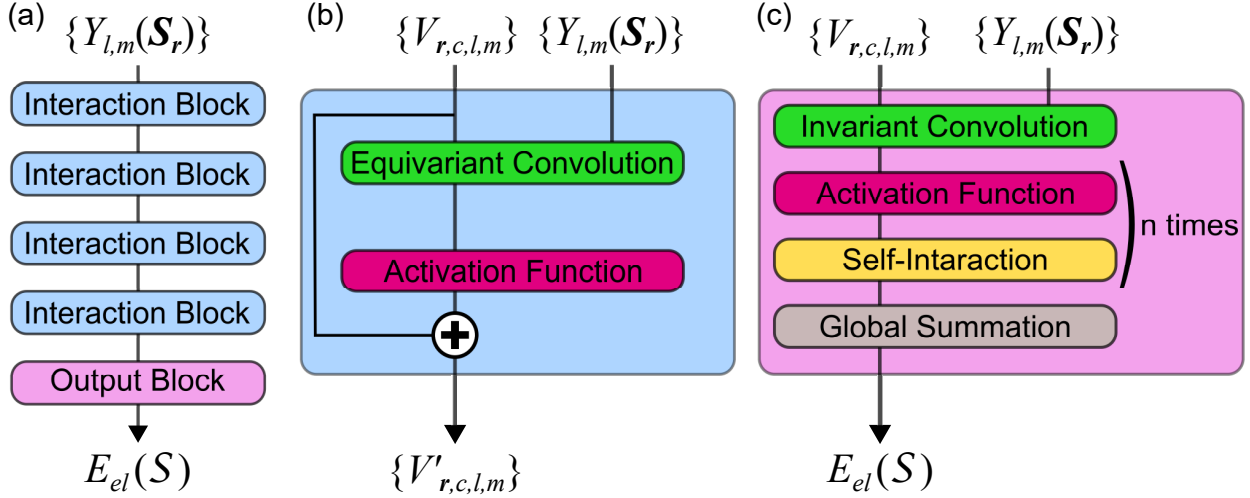}
    \caption{Architecture of (a) whole equivariant convolutional neural networks, (b) interaction block, and (c) output block.}
    \label{fig:arch}
\end{figure*}

In condensed matter physics, many physical properties are described by tensors.
Their tensor properties are dominated by the symmetries of the physical system.
Equivariance can more generally represent the tensor properties and the tensor operations (e.g. vector addition, dot products, and cross products) \cite{Batzner2022,Thomas2018}.
A function $f: \mathcal{X}\to\mathcal{Y}$ (for vector spaces $\mathcal{X}$ and $\mathcal{Y}$) is \textit{equivariant} with respect to a group $G$ and group representations $\mathcal{D}_\mathcal{X}(g)$ and $\mathcal{D}_\mathcal{Y}(g)$ if for all $g \in G$,
\[f \circ \mathcal{D}_\mathcal{X}(g) = \mathcal{D}_\mathcal{Y}(g) \circ f. \]
I note that \textit{invariance} is a specific case of $\mathcal{D}_\mathcal{Y} = \mathrm{id}_\mathcal{Y}$. 
In other words, equivariance is an extension of the concept of invariance.

In dealing with classical spins, it is necessary to discuss equivariance with respect to $SO(3)$ \cite{Thomas2018}. Spherical harmonics $Y_{l,m}(\hat{\bm{r}})$ (where $l=0,1,\ldots$ is the degree, $m=-l,-l+1,\ldots,l$ is the order, and $\hat{\bm{r}}$ is a unit vector) play an important role. This is because $Y_{l,m}(\hat{\bm{r}})$ is equivariant to $SO(3)$. In other words, for any $g \in SO(3)$, 
$$Y_{l,m}(\mathcal{R}(g)\hat{r})=\sum_{m^{\prime}}D_{l,mm^{\prime}}(g)Y_{l,m^{\prime}}(\hat{r})$$ 
holds.
Here, $\mathcal{R}$ are the rotation matrices and $D_{l,mm^{\prime}}$ are the Wigner D-matrices \cite{Gilmore2008}.

What is even more important is that the tensor product of irreducible representations of $SO(3)$ can be decomposed into new irreducible representations using Clebsch-Gordan coefficients. Specifically, the tensor product of two irreducible representations $u_{l_1}$ of degree $l_1$ and $v_{l_2}$ of degree $l_2$ is decomposed into a direct sum of irreducible representations of degrees from $l_1 + l_2$ to $|l_1 - l_2|$. This decomposition is expressed as:
$$(u\otimes v)_{l,m} = \sum_{m_1,m_2}C^{l,m}_{l_1,m_1,l_2,m_2}u_{l_1,m_1}v_{l_2,m_2}.$$
Here, $C^{l,m}_{l_1,m_1,l_2,m_2} = \braket{l,m|l_1,m_1,l_2,m_2}$ are the Clebsch-Gordan coefﬁcients.
For the sake of simplifying notation, this operation is sometimes written as $l_1 \otimes l_2 \to l$.
The tensor product is also equivariant. 
Interestingly, the tensor products include fundamental vector operations. 
For example, $1 \otimes 1 \to 0$ and $1 \otimes 1 \to 1$ correspond to the dot and cross products of vectors, respectively \cite{Thomas2018}. 
This fact suggests that the tensor product is well-suited for describing classical spin models.

\begin{figure*}
    \centering
    \includegraphics[width=\textwidth]{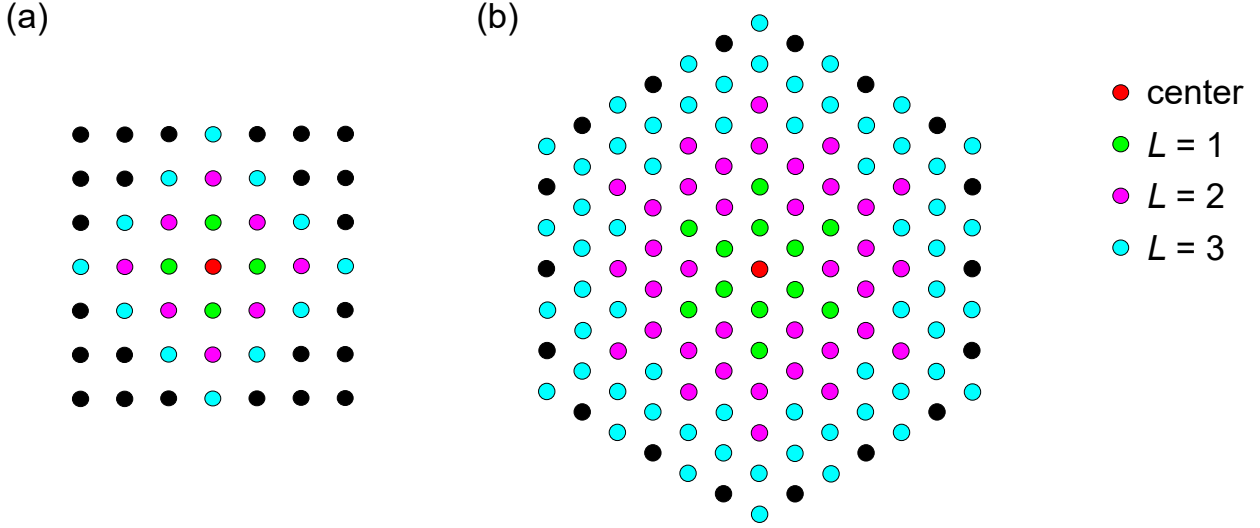}
    \caption{Local spin environment in an ECNN with $L$ convolution layers. (a) Square lattice with only nearest-neighbor hopping, (b) Triangular lattice with both nearest-neighbor and third-neighbor hopping.}
    \label{fig:LSE}
\end{figure*}

\subsection{Equivariant Neural Networks Potential for Spin Configurations}
To simulate localized spin configurations (e.g. Monte Carlo sampling, simulated annealing, and Landau-Lifshitz-Gilbert (LLG) simulation), the energy $E(\mathcal{S})$ or the effective magnetic fields $\bm{B}_{\bm{r}}(\mathcal{S}) = - \frac{\partial E}{\partial \bm{S}_{\bm{r}}}$ are usually needed.
For concreteness, I focus on the (stochastic) LLG simulation.
In the adiabatic limit, the time evolution of spins at finite temperature is governed by the stochastic LLG equation 
\begin{equation}
    \frac{d\bm{S}_{\bm{r}}}{dt} = - \bm{S}_{\bm{r}} \times \left(\bm{B}_{\bm{r}}(\mathcal{S})+\bm{\zeta}_{\bm{r}}\right) + \alpha \bm{S}_{\bm{r}} \times \frac{d\bm{S}_{\bm{r}}}{dt}.
\end{equation}
Here, $\bm{\zeta}_{\bm{r}}$ is the Gaussian stochastic fields representing thermal fluctuations and $\alpha$ is the Gilbert damping constant.
The energy $E(\mathcal{S})$ can be divided into $E = E_{el} + E_s$ as same as eq.~\eqref{eq:H}.
The classical spin energy $E_s=\mathcal{H}_s$ can be easily computed, however, the itinerant electron energy $E_{el}=\braket{\mathcal{H}_{el}}$ needs exact diagonalizations of $\mathcal{H}_{el}$, whose computational cost is ordinarily ${O}(N^3)$.
Here, $\braket\cdot$ is the expectation value for itinerant electrons.

In the NN model, I assume locality for the total energy of the itinerant electrons, $E_{el}(\mathcal{S})$, and decompose it into local contributions \cite{Zhang2021} as $$E_{el}(\mathcal{S})=\sum_{\bm{r}}\epsilon(\mathcal{S}_{\bm{r}}).$$ 
Here, $\mathcal{S}_{\bm{r}}$ is the set of spins considered as the local environment at position $\bm{r}$ (call $\mathcal{S}_{\bm{r}}$ a local spin environment (LSE) later), and in the NN model, it is determined by the kernel $\mathcal{N}(\bm{r})$ and the depth of the layers. 
The NN model predicts this local energy $\epsilon(\mathcal{S}_{\bm{r}})$, and this assumption allows the NN model to use different lattice sizes during training and simulation.
As for the effective magnetic field, it can be computed efficiently using automatic differentiation.

Figure~\ref{fig:arch}(a) illustrates the configuration of the neural network used in this study, which consists of stacked interaction blocks and a final output block. Interaction blocks (Fig.~\ref{fig:arch}(b)), which extract equivariant features are composed of equivariant convolution layers and nonlinear layers. 
For stability of training processes, ResNet type skip-connections \cite{He2016} are adopted.
The output block (Fig.~\ref{fig:arch}(c)) is made up of a convolution layer with degree $l=0$ output, nonlinear layers, and  self-interaction layers, finally producing an energy output that is invariant to translation and rotation operations.

\subsubsection{equivariant convolution layer}
The equivariant convolution layers $\mathcal{L}$ are given by
\begin{widetext}
    \begin{equation}\label{eq:EC}
         \widetilde{V}_{\bm{r},c_o,l_o,m_o} = \mathcal{L}_{\bm{r},c_o,l_o,m_o}(\mathcal{V}) = \sum_{l_f,l_i,m_f,m_i,c_i}\sum_{\bm{r}'\in\mathcal{N}(\bm{r})}W_{\bm{r}'-\bm{r},c_i,l_i,l_f}^{c_o,l_o}C^{l_o,m_o}_{l_i,m_i,l_f,m_f} Y_{l_f,m_f}(\bm{S}_{\bm{r}})V_{\bm{r}',c_i,l_i,m_i}.
    \end{equation}
\end{widetext}
Here, $V_{\bm{r},c,l,m}$ and $\widetilde{V}_{\bm{r},c,l,m}$ are feature vectors of inputs/outputs of the layer $\mathcal{L}$ at the position $\bm{r}$, the channel $c$, the degree $l$, and the order $m$. $\mathcal{V} = \{V_{\bm{r},c,l,m}\}$  is a set of feature vectors. 
The subscriptions such as $i$, $f$, and $o$ describe ``input'', ``filter'', and ``output'', respectively.
$W_{\bm{r}'-\bm{r},c_i,l_i,l_f}^{c_o,l_o}$ are neural network parameters of equivariant convolutions,  $Y_{l,m}$ are the spherical harmonics, and $\mathcal{N}(\bm{r})$ is the set of positions around position $\bm{r}$.
In usual CNNs, the main layers consist of filters (also called kernels), which slide over the image (or the feature map), performing element-wise multiplication and summing the results.  
This operation guarantees translational equivariance \cite{Cohen2016}.
The equivariant convolution in this paper guarantees rotational equivariance for the spins and translational equivariance for the lattice by replacing element-wise multiplication with the tensor product involving the spherical harmonics of the central spin.
In this paper, $\mathcal{N}(\bm{r})$ is defined as the set of positions connected to the center $\bm{r}$ itself and those connected to $\bm{r}$ through direct hopping terms, i.e., $\mathcal{N}(\bm{r})=\left\{\bm{r}'\middle|\bm{r}'=\bm{r}\ \mathrm{or}\ t_{\bm{r}\bm{r}'}\neq0\right\}$.
Figure~\ref{fig:LSE} shows LSEs for the case of a square lattice with nearest-neighbor hopping and a triangular lattice with both nearest-neighbor and third-nearest-neighbor hopping.
I use e3nn library \cite{e3nn,e3nn_paper} to implement the tensor product, which is based on PyTorch \cite{paszke2019pytorch}.

\subsubsection{activation function}
In usual neural networks, nonlinear functions called activation functions $\eta(x)$ are applied to each component of the feature vector $\bm{x}$.
However, our model requires the following transformation in order to maintain the equivariance:
\begin{equation}\label{eq:activation}
    \begin{cases}
        \eta\big{(}V_{\bm{r},c,0,0}\big{)} & (l = 0) \\
        \eta\big{(}\|V\|_{\bm{r},c,l}+b_{c,l}\big{)} {V_{\bm{r},c,l,m}}/{\|V\|_{\bm{r},c,l}} & (l \geq 1)      
    \end{cases}
\end{equation}
for $\|V\|_{\bm{r},c,l}:=\sqrt{\sum_{m}|V_{\bm{r},c,l,m}|^{2}}$ \cite{Thomas2018}.
I use the swish activation function \cite{Hendrycks2016,Elfwing2018}
\begin{equation}
    \eta(x) = \frac{x}{1+e^{-x}}
\end{equation}
for differentiability in this paper. 

\subsubsection{self-interaction}
I use point convolutions
\begin{equation}
    \sum_{c}W_{lc}^{c'}V_{\bm{r},c,l,m},
\end{equation}
which scale the feature vectors elementwise and mix the components of the feature vectors at each point \cite{Schutt2017,Thomas2018}.
To maintain equivariance, the same weights should be used for every $m$.

\subsection{Performance in Square Lattice}
To evaluate the efficiency, I perform training and prediction using the ECNN model in a square Kondo lattice model. 
The Kondo lattice model assumes nearest-neighbor hopping $t_1=1.0$, the Hund coupling parameter $J=7.0$, and electron filling fraction $n=0.485$. Data is obtained by performing exact diagonalization for 100 random spin configurations on a $30\times 30$ lattice and using the resulting effective magnetic field for training. 
Additionally, 50 independent validation data samples are prepared.
See Method for details for the details of training conditions.
\begin{figure*}
    \centering
    \includegraphics[width=\textwidth]{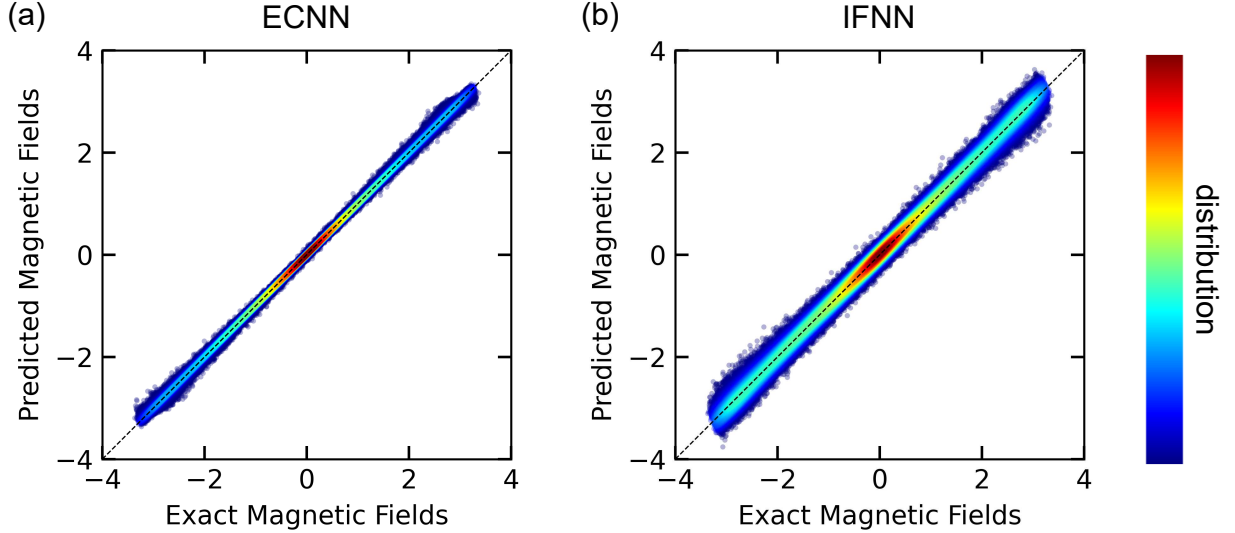}
    \caption{Effective magnetic field calculated by ED versus effective magnetic field predicted by (a)~equivariant convolutional neural network (ECNN) and (b)~the invariant-descriptor-based fully connected neural network (IFNN). The colors represent the data distribution obtained by the kernel density estimation.}
    \label{fig:sq_pred}
\end{figure*}

Figure~\ref{fig:sq_pred}(a) shows components of the effective magnetic field predicted by the ECNN model after training and the exact results obtained through exact diagonalization (ED).
Based on the RMSE criterion, the training and validation errors are $0.0466$ and $0.0481$, respectively, indicating surprisingly high accuracy in the predictions, despite using only 100 data points. 

For comparison, Fig.~\ref{fig:sq_pred}(b) presents the results of prediction with the same data using the invariant-descriptor-based fully connected neural network (IFNN) in ref. \citealp{Zhang2021}. 
The code and conditions used for training are taken from the Github repository \cite{Zhang2021}.
Based on the RMSE criterion, the training and validation errors are $0.124$ and $0.148$, respectively, with the validation error being more than three times larger for the ECNN. It is worth noting that in the IFNN, the validation error slightly increases after reaching its minimum value in a few epochs, while the training error continues to decrease, exhibiting typical overfitting dynamics. The final training and validation errors are $0.109$ and $0.150$, respectively. In contrast, the difference between the training and validation errors for the ECNN is minimal. This suggests that the inductive bias of the ECNN model is much stronger than that of the IFNN model.

To investigate the scalability of the ECNN model, I examine the computation time when varying the lattice size of the input spin configurations. 
Both the ECNN predictions and the ED  calculations are performed on a single NVIDIA Tesla A100 (40 GB) GPU. 
Figure~S1 shows the average computation time for ECNN predictions and ED calculations over 10 predictions. For small lattice sizes, the ECNN prediction time shows almost no size dependency, while for larger lattice sizes, it generally exhibits linear scaling. 
Even for a very large lattice size of $576\times576$, energy and effective magnetic field can be calculated in just about one second. 
When comparing the ECNN prediction time to the ED calculation time, there is approximately a 700-fold speed-up in the case of $128\times 128$, which is the limit lattice size that can be calculated using ED. 
Although lattice sizes larger than $576\times 576$ could not be calculated in this study due to GPU memory limitations, it is possible to handle larger sizes using multiple GPUs by dividing the problem based on the assumption of locality.
It should be noted that when evaluating the prediction error for all sizes that are computationally feasible with ED using the ECNN model trained on $32\times 32$ data, I do not observe a worsening of prediction error due to the change in lattice size.

\subsection{Dynamics Simulation in Triangle Lattice}
\begin{figure*}
    \centering
    \includegraphics[width=\textwidth]{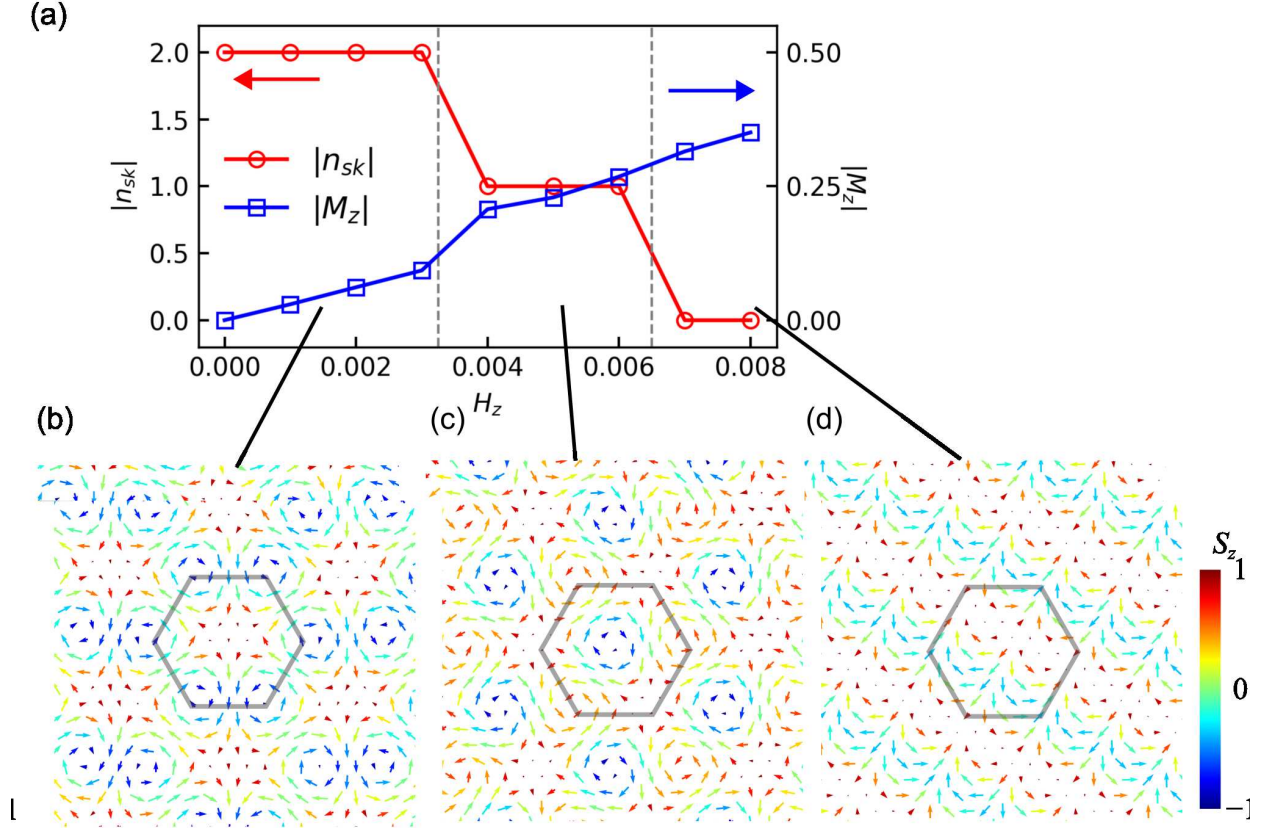}
    \caption{(a)~$H_z$ dependences of the topological number $|n_{sk}|$ and the magnetization $|M_z|$ obtained by the
ML-LLG simulation. (b)\textendash(d)~Configurations of localized spins in (b) the $|n_{sk}|=2$ SkX at $H_z=0.002$, (b) the $|n_{sk}|=1$ SkX at $H_z=0.005$, and (c) the $|n_{sk}|=0$ state at $H_z=0.008$. The gray hexagons in (b)\textendash(d) represent the magnetic unit cell.}
    \label{fig:sk}
\end{figure*}
To investigate whether more complex magnetic structures can be reproduced, I examine the dynamics on the triangular lattice. 
In previous work \cite{Hayami2021locking,Hayami2021temperature,Eto2022,Ozawa2017}, skyrmion crystal (SkX) phases have been extensively studied in the system with first-nearest-neighbor hopping $t_1=1.0$, third-nearest-neighbor hopping $t_3=-0.85$, Hund coupling $J=-1.0$, and chemical potential $\mu=-3.5$. 
Training data is generated for the $36\times 36$ site triangular lattice Kondo lattice model with temperature $T=0.005$ and Gilbert damping constant $\alpha=1.0$ by performing stochastic ED-LLG simulations with three different magnetic fields $H_z=0.002, 0.005, 0.008$. 
I sample 100 data points each from non-equilibrium and equilibrium states for each magnetic field and add 300 random spin configuration data points for a total of 900 data points. 
The validation data consists of 100 samples each from the equilibrium states for each magnetic field, totaling 300 samples.
I note that in the $36\times 36$ ED-LLG simulations, I'm not able to observe any skyrmions (typical snapshots are shown in Fig.~S2).
The stability of skyrmions is heavily influenced by the lattice's rotational symmetry. 
To ensure the rotational symmetry,  data augmentation, which uses data with the lattice randomly rotated by $n\pi/3$ during training (for $n=0,1,\ldots,5$), is performed.
After training, the ECNN model's RMSE-based training and validation errors for the effective magnetic field are 0.00121 and 0.00104, respectively, while for the energy density, they are 0.000281 and 0.0000415, respectively.

I perform machine learning LLG (ML-LLG) simulations at $T=0$ using the trained ECNN model on a $96\times96$ site triangular lattice. According to ref. \citealp{Ozawa2017,Eto2022}, a transition from the bi-skyrmion phase ($n_{sk}=2$) to the skyrmion phase ($n_{sk}=1$) occurs at  $H_z\sim0.00325$, and a transition from the skyrmion phase ($n_{sk}=1$) to the non-topological phase ($n_{sk}=0$) occurs at $H_z\sim0.0065$. In Fig.~\ref{fig:sk}(a), I present the topological number $n_{sk}$ and localized spin magnetization per site $|M_z|$ obtained from the ML-LLG simulations. The ECNN model successfully reproduces the phase transitions of the bi-skyrmion phase, the skyrmion phase, and the non-topological phase. Moreover, I show the typical spin configurations of each phase obtained from the ML-LLG simulations in Figs.~\ref{fig:sk}(b)\textendash(d). The spin configurations closely resemble those from ref. \citealp{Ozawa2017,Eto2022}, demonstrating that the ECNN model can accurately reproduce complex orders such as skyrmion crystals.

\section{Discussion}
I consider the reasons why the ECNN model demonstrates excellent predictive accuracy. The first reason is that equivariant operations with respect to symmetry transformations tend to preserve information more easily. This point has also been noted in Ref. \citealp{Cohen2016}, which first drew attention to equivariant neural networks. On the other hand, when using invariant descriptors, a significant loss of information occurs at the point of transformation to the descriptors.
From the  point of view of physics, for example, when evaluating the energy of a molecule, it is expressed as a simple mathematical operation of taking the expectation value of the Hamiltonian with its eigen wavefunction. This eigen wavefunction is equivariant with respect to the symmetry operations of the Hamiltonian as a function of coordinates, but it is not invariant. Although Hohenberg-Kohn's series of theories does not forbid deriving energy from the invariant electron density, the fact that the specific form of the functional is not known to this day suggests that it is not a simple functional. From this analogy, using equivariant layers may offer the possibility of achieving physically correct solutions "more simply" than using invariant descriptors.

The second reason is that the convolution operation reflects the graph topology of the hopping in the Kondo lattice model. In the case of a fully connected neural network, the treatment of spin combinations within a given cutoff is equivalent. On the other hand, in convolutional neural networks, the LSE is expanded by stacking kernels that reflect the hopping connections (Fig.~\ref{fig:LSE}). The graph topology is reflected, with the spin of the directly connected sites to the center being the most important, and the spin of sites with a greater number of intervening connections being less emphasized.
Figure~S3 shows the correlation between the spin and the effective magnetic field in the skyrmion crystal phase for both ED and ECNN model cases. Indeed, the correlation in the ECNN model reproduces that of the ED well, indicating that the graph topology represented by the ECNN model provides a physically valid description.

In the application of deep learning in materials science, one of challenges is the preparation of  a large amount of high-quality data. 
If the cost of collecting the data needed for training is high, deep learning will not be practical, even if its performance is superior.
Similar to other equivariant neural networks \cite{Batzner2022}, the ECNN model can make highly accurate predictions with a small amount of data and small lattice sizes. 
In tasks that require a broad parameter space exploration, such as creating phase diagrams or optimizing material properties, this efficiency is clearly advantageous.

Furthermore, it can be considered that the intermediate layers of the ECNN model generate feature vectors suitable for representing the quantum states of itinerant electrons, as the tensor products can represent general vector and tensor operations in physics and can accurately evaluate energies. Therefore, by performing transfer learning or fine tuning, it could be possible that the ECNN model is used to predict not only energy and effective magnetic fields but also other quantum properties such as optical properties and  transport properties.

\section{Conclusion}
In this paper, I develop an equivariant convolutional neural network architecture that accelerates spin dynamics in systems where itinerant electrons and localized spins are interacted. The tensor-product-based convolution ensures equivariance with respect to spin rotation and lattice translation. I implement and verify this approach for both square and triangular lattices.
For the square lattice, the developed method exhibits higher accuracy than invariant descriptor-based neural networks. Furthermore, it can perform large-scale calculations with $572 \times 572$ sites in just about 1 second. In the case of the triangular lattice, it is found to have sufficient accuracy for evaluating phase transitions in the skyrmion crystal phases.

\begin{acknowledgments}
I would like to thank Yuki Shiomi and Tomoyuki Yokouchi for useful comments.
This work is supported by Grant No.~21J20969.
Some calculations were conducted using the  FUJITSU Server PRIMERGY GX2570 (Wisteria/BDEC-01) at the Information Technology Center, The University of Tokyo.
Y.M. is supported by Research Fellowships of Japan Society for the Promotion of Science for Young Scientists.
\end{acknowledgments}

\section{Method}
\subsection{Neural Network Models and Training Details}
As mentioned in the main text, the architecture of the equivariant convolutional neural network consists of four interactive blocks and an output block. Within the interactive block, when the degree of the input feature vector within the filter is $l_i$, the degree of the spherical harmonic function of the central spin used as a filter is $l_f$, and the degree of the output feature vector is $l_o$, the tensor product of $l_i\otimes l_f\to l_o$ is performed, followed by applying the activation function to each site.
In the interaction block, I set $l_i=\{0,1,2\}$, $l_f=\{0,1\}$, and $l_o=\{0,1,2\}$, with the number of channels being 8. In the first layer, spherical harmonic functions of spins are used as the feature vector. In the output block, first, the tensor product is set to $l_i\otimes l_f\to 0$ (invariant convolution), and then the self-interaction, which changes  the number of channels for each site, and the activation function are performed repeatedly. The number of channels for self-interaction changes as $8\to 4\to 2\to 1$.

During the training, I use the sum of the mean squared errors (MSE) of the effective magnetic field and the energy density as the loss function: 
$$\frac{1}{(3L_xL_y)^2}\sum_{\bm{r}}\left|\bm{B}_{\bm{r}}-\bm{B}^*_{\bm{r}}\right|^2 + \frac{\lambda}{(L_xL_y)^2}\left|E-E^*\right|^2.$$ 
Here, $(L_x, L_y)$ represents the lattice size, and $\bm{B}_{\bm{r}}^*$ and $E^*$ are the effective magnetic field and energy calculated by exact diagonalization. For the square lattice, I prioritize the accuracy of the effective magnetic field by setting $\lambda=0.0$, and for the triangular lattice, I evaluate both energy and the effective magnetic field by setting $\lambda=1.0$.
As an optimizer, I adopt AdaGrad \cite{Duchi2011}, and if the error does not improve during 20 epochs, the learning rate is halved.
All models are trained on a NVIDIA Tesla A100 (40 GB) GPU in single-GPU training using ﬂoat32 precision.
\section{Data Availability}
The software developed in this study is released under the MIT License and publicly available. Interested parties can access the source code and documentation at the following GitHub repository: \url{https://github.com/Miyazaki-Yu/ecnn4klm}.
\bibliography{ref}



\end{document}



\title{Supplemental Materials for: \\ Equivariant Neural Networks for Spin Dynamics Simulations of Itinerant Magnets}

\author{Yu Miyazaki}
\email[]{yumiyazaki@g.ecc.u-tokyo.ac.jp}
\affiliation{Department of Applied Physics, The University of Tokyo, Hongo, Bunkyo, Tokyo 113-8656, Japan}

\date{\today}
\maketitle

\section{Correlation Test}
In periodic structures such as skyrmion crystal phases, the distance dependence of the correlation between spins and effective magnetic fields is crucial. Therefore, I investigate the correlation of the effective magnetic field in the triangular lattice model following reference \cite{Zhang2021}. First, I randomly select a spin $\bm{S}_{\bm{r}}$ from the spin configuration $\mathcal{S}$ and rotate it randomly. I denote the new spin configuration as $\mathcal{S}'$, and calculate the change in the effective magnetic field $\Delta\bm{B}=\bm{B}(\mathcal{S}')-\bm{B}(\mathcal{S})$. I then plot the magnitude of the change normalized by the original effective magnetic field $|\Delta\bm{B}/\bm{B}|=|\bm{B}(\mathcal{S}')-\bm{B}(\mathcal{S})| /|\bm{B}(\mathcal{S})|$ as a function of the distance $r$ from the center.

\begin{figure*}
    \centering
    \includegraphics[width=\textwidth]{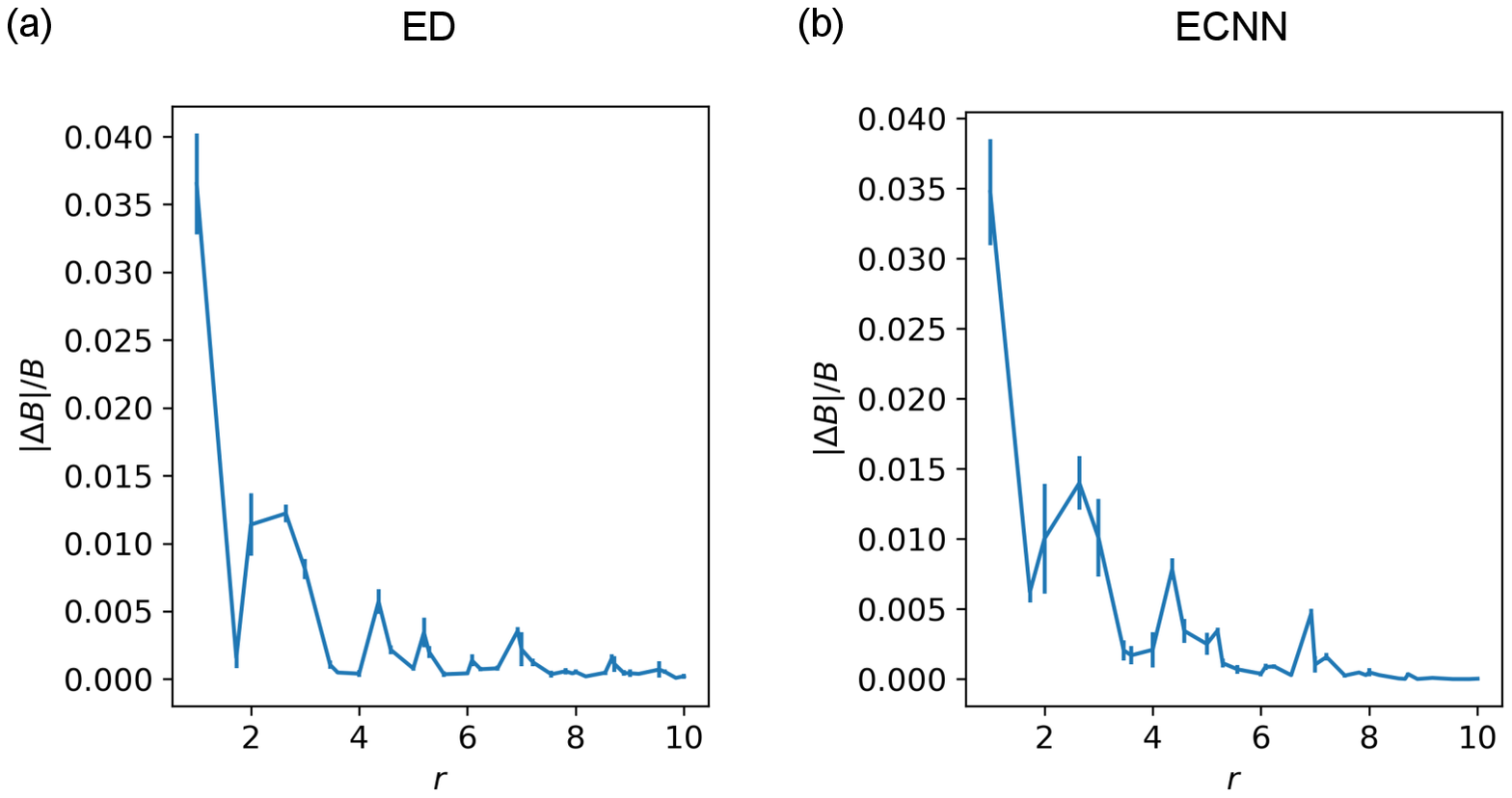}
    \caption{Distance dependence of the correlation between spin and effective magnetic field. (a) Exact diagonalization (ED). (b) Equivariant convolutional neural network (ECNN)}
    \label{fig:B_diff}
\end{figure*}

Figure~\ref{fig:B_diff}(a) shows the results for this operation using exact diagonalization (ED), and Figure~\ref{fig:B_diff}(b) shows the results using the equivariant convolutional neural network (ECNN). The model in this study has two hopping terms, the first and third nearest neighbors, resulting in a complex distance dependence. Nevertheless, the ECNN generally reproduces the peak positions. This demonstrates that the convolution in the ECNN can represent the graph structure of the Kondo lattice model. Furthermore, for $r\gtrsim 7$, there is almost no correlation between the spins and the effective magnetic field, indicating that the assumption of locality is valid.
\begin{figure*}
    \centering
    \includegraphics[width=\textwidth]{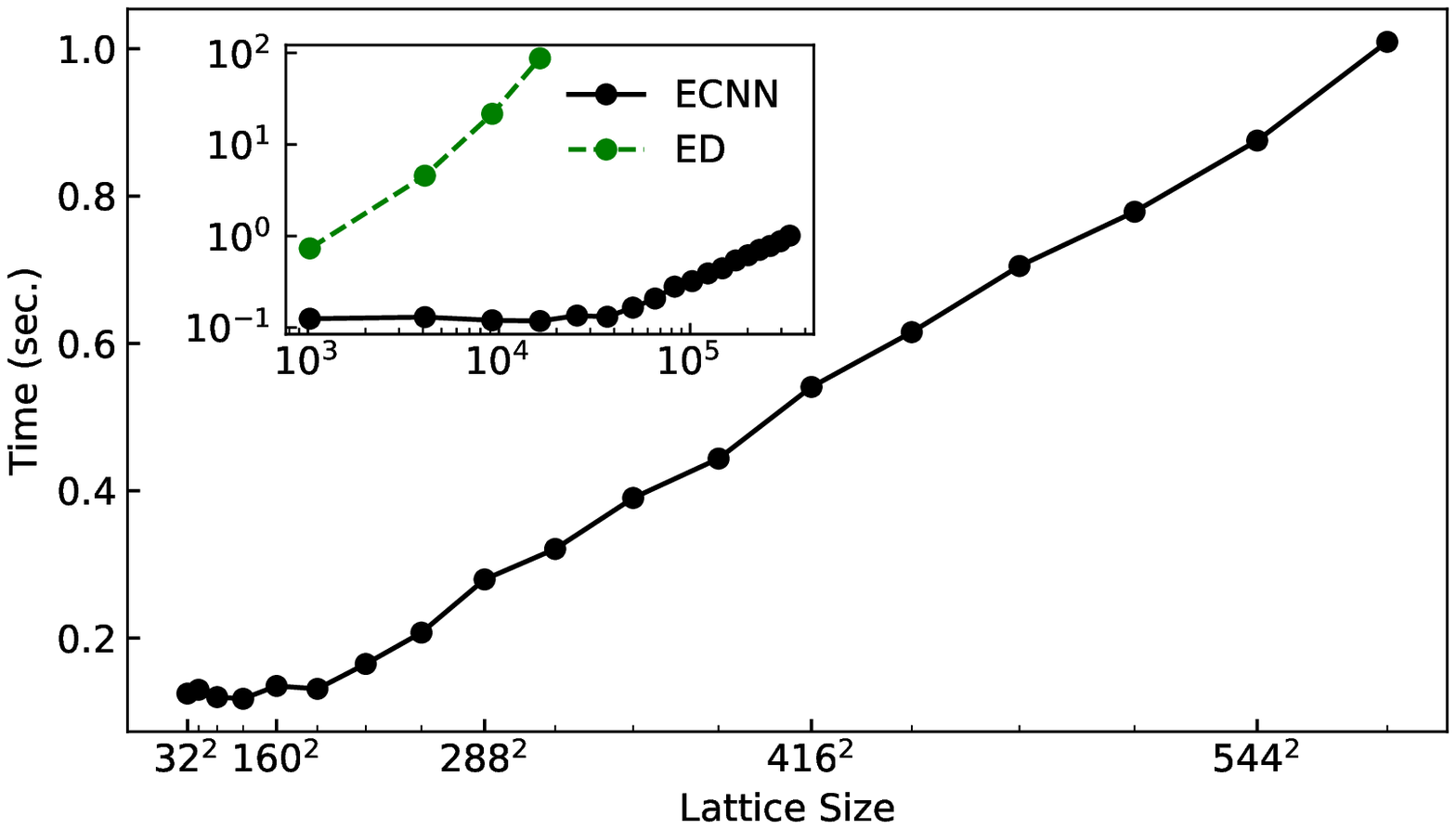}
    \caption{Calculation time dependence on the lattice size in square lattices. The black solid line represents the prediction time of the ECNN, while the green dashed line represents the calculation time of exact diagonalization (ED). The main window displays a linear plot, and the inset shows a log-log plot. Both calculations are performed on an NVIDIA Tesla A100 GPU (40 GB).}
    \label{fig:time}
\end{figure*}

\begin{figure*}
    \centering
    \includegraphics[width=\textwidth]{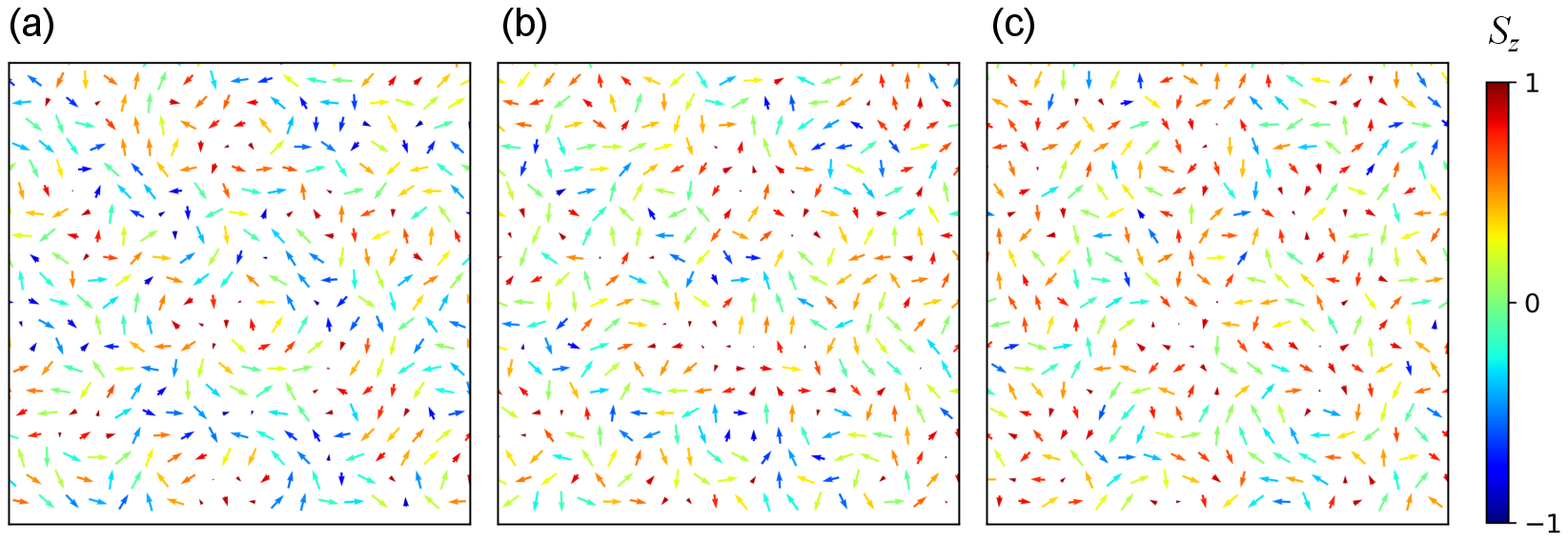}
    \caption{Snapshots of exact diagonalization LLG simulations for a $36\times36$ site triangular lattice Kondo model. (a) $H_z=0.002$, (b) $H_z=0.005$, and (c) $H_z=0.008$.}
    \label{fig:ED}
\end{figure*}

\bibliography{ref}